\documentclass[prb,aps,twocolumn,reprint,amsmath,amssymb,showpacs,superscriptaddress]{revtex4-1}

\usepackage{bm}
\usepackage{dcolumn}
\usepackage{graphicx}
\usepackage{amsfonts}
\usepackage{amsmath}
\usepackage{amssymb}
\usepackage{subfigure}
\usepackage{color}
\newcommand{\beq}{\begin{equation}}
\newcommand{\eeq}{\end{equation}}
\newcommand{\beqarray}{\begin{eqnarray}}
\newcommand{\eeqarray}{\end{eqnarray}}
\newcolumntype{.}{D{.}{.}{-1}}

\allowdisplaybreaks

\begin{document}
\title{Spectral density in a nematic state of iron pnictides}

\author{Maria Daghofer}
\email{M.Daghofer@ifw-dresden.de} 
\affiliation{IFW Dresden, P.O. Box 27 01 16, D-01171 Dresden, Germany}
\author{Andrew Nicholson}
\author{Adriana Moreo}
\affiliation{Department of Physics and Astronomy, The University of
  Tennessee, Knoxville, TN 37996} 
\affiliation{Materials Science and Technology Division, Oak Ridge
  National Laboratory, Oak Ridge, TN 32831}

\date{\today}

\begin{abstract}
Using cluster-perturbation theory, we calculate the
spectral density $A({\bf k},\omega)$ for a nematic phase of models
describing pnictide superconductors, where very short-range 
magnetic correlations choose the ordering vector $(\pi,0)$ over the
equivalent $(0,\pi)$ and thus, break the fourfold rotation symmetry of
the underlying lattice without inducing long-range magnetic order. 
In excellent agreement with angle-resolved photo-emission spectroscopy
(ARPES), we find that the $yz$ bands at $X$ move to higher
energies. When onsite Coulomb repulsion brings the system close to a
spin--density-wave (SDW) and renormalizes the band width by a factor
of $\approx 2$, even small anisotropic couplings of $10$ to
$15\;\textrm{meV}$ strongly
distort the bands, splitting the formerly degenerate states at $X$ and $Y$
by $\approx 70\;\textrm{meV}$ and shifting the $yz$ states at $X$ above the chemical 
potential. This similarity to the SDW bands is in excellent agreement
with ARPES. An important difference to the SDW bands
  is that the $yz$ bands still cross the Fermi level, again in
  agreement with experiment. 
We find that orbital weights near the Fermi surface provide a better
characterization than overall orbital densities and orbital polarization.
\end{abstract}

\maketitle

\section{Introduction} 


In recent years, iron-based superconductors have been intensely
studied,~\cite{pnict_rev_johnston,Paglione:2010p2493} because of
 their high superconducting transition temperatures. As in the
cuprates, antiferromagnetic (AFM) order is present in the phase
diagram and phonons are not believed to be strong enough to explain the
high transition temperatures.~\cite{phonon0} 
However, in the pnictides the AFM phase is a metallic spin-density wave (SDW) 
rather than a system of localized Heisenberg spins as is the case in the cuprates.
At temperatures slightly above the
transition to the SDW with ordering vector $(\pi,0)$ or $(0,\pi)$,
many weakly doped compounds show an orthorhombic phase without
long-range magnetic order, but with broken rotational symmetry. This
phase has slightly different lattice constants along the in-plane 
iron-iron bonds,~\cite{neutrons1} but the anisotropy that develops in  
electronic observables such as
resistivity~\cite{rho_anisotropy,Anis_charge} or angle-resolved
photoemission spectroscopy (ARPES)~\cite{Yi:PNAS2011,ARPES_NaFeAs11,He:2010pNaFeAs,ZhangARPES_NaFeAs_2011}
of detwinned samples appears considerably more pronounced.  

A number of competing scenarios have been proposed for this phase and can be broadly
categorized as ``magnetism'', ``orbital'', and ``lattice'' driven. In
the first case, the symmetry between equivalent
magnetic ordering vectors $(\pi,0)$ and $(0,\pi)$ is broken and the
system chooses one of them without immediately establishing long-range
magnetic order.~\cite{Fang:2008nematic,Xu2} In the second picture, it
is the degeneracy between two $d$ orbitals of the iron ion, the $xz$ and
$yz$ states providing the greatest contribution to the states at the Fermi surface (FS),
that is spontaneously broken;~\cite{kruger:054504} the resulting
orbital occupation then determines the effective magnetic exchange constants 
that generate the SDW order. Both pictures were first discussed
in insulating spin and spin-orbital models and have since been
generalized to take into account electron
itineracy. Studies in several models have shown that 
nematic phases can indeed develop between structural and magnetic
transition temperatures.~\cite{Fernandes_nematic2012,PhysRevB.82.020408}

While a definite
answer about the driving mechanism may be hard to nail down, as spin,
orbital,~\cite{kruger:054504,Fernandes_nematic2012} and
lattice~\cite{PhysRevB.79.180504,Paul:2011p2617,PhysRevB.82.020408} degrees of freedom
are naturally coupled and 
interact with each other, one may nevertheless try to identify
the dominant ingredient(s). To this end, it is instructive to 
establish how each type of symmetry breaking manifests itself in
observables. If the symmetry breaking is assumed to mostly concern the
$xz$ and $yz$ orbitals, one can introduce it explicitly by adding 
a phenomenological energy splitting between the orbitals and evaluating
its impact on observables such as the optical conductivity or the
spectral density. These signatures were found to qualitatively agree
with experiments,\cite{oo_nematic_2011} where states 
of $yz$ character are found to be higher in energy than those of $xz$
character in several different pnictide compounds from 
the two structurally slightly different``111'' and ``122'' families. On the other hand, ARPES data taken above the N\'eel
temperature have alternatively been interpreted in terms of ``band folding''
due to magnetic order~\cite{He:2010pNaFeAs} or emphasising 
the coupling between magnetic order and the orbital states near the
Fermi level.~\cite{ZhangARPES_NaFeAs_2011} 
Short-range magnetic order~\cite{Shuhua_Pnict2011} and the spin-nematic scenario~\cite{Fernandes:2011transp} likewise
reproduce the anisotropic conductivity,
and the latter has been argued to lead to an effective orbital
splitting.~\cite{Fernandes_nematic2012} However, a direct calculation of
the spectral density in a nematic phase is so far lacking.

We use here a method that combines real and momentum space,
cluster-perturbation theory
(CPT),~\cite{Gros:1993p2667,Senechal:2000p2666} to calculate
the one-particle spectral density $A({\bf k},\omega)$ for 
a spin-nematic phase where rotational symmetry is broken via (very)
short-range spin correlations that are AFM in $x$ and
ferromagnetic (FM) in $y$ direction [corresponding to ordering vector $(\pi,0)$], but
without long-range magnetic correlations beyond second neighbors. 
The obtained spectral density reproduces the momentum-dependence of
the band shifting observed in
ARPES.~\cite{ARPES_NaFeAs11,He:2010pNaFeAs,ZhangARPES_NaFeAs_2011}
If onsite interactions bring the
system close to the SDW transition, a small phenomenological magnetic
anisotropy leads to large anisotropies in $A({\bf k},\omega)$. Thus, we can
theoretically describe the astonishing ARPES result that the
overall band shifts  characterizing $A({\bf k},\omega)$ in the SDW
phase are nearly fully developed already above the N\'eel
temperature.~\cite{ARPES_NaFeAs11}

In
order to be able to solve the problem exactly on a small cluster, we
use variants of models with three~\cite{Daghofer_3orb} and four~\cite{PhysRevB.84.235115}
orbitals; models and method are
introduced and discussed in
Sec.~\ref{sec:modelmethod}. Section~\ref{sec:results_U0} discusses the
anisotropic band shifts induced by (strong) anisotropic magnetic couplings in the
non-interacting models; in Sec.~\ref{sec:results_U}, we show that in
the presence of onsite interactions and near the SDW transition,
smaller magnetic anisotropies have a large impact. In
Sec.~\ref{sec:conclusions}, our results are summarized and discussed.

\section{Method and Model}\label{sec:modelmethod}

\begin{figure}
\begin{minipage}{0.33\columnwidth}
\subfigure[]{\includegraphics[width=\textwidth,angle=90]{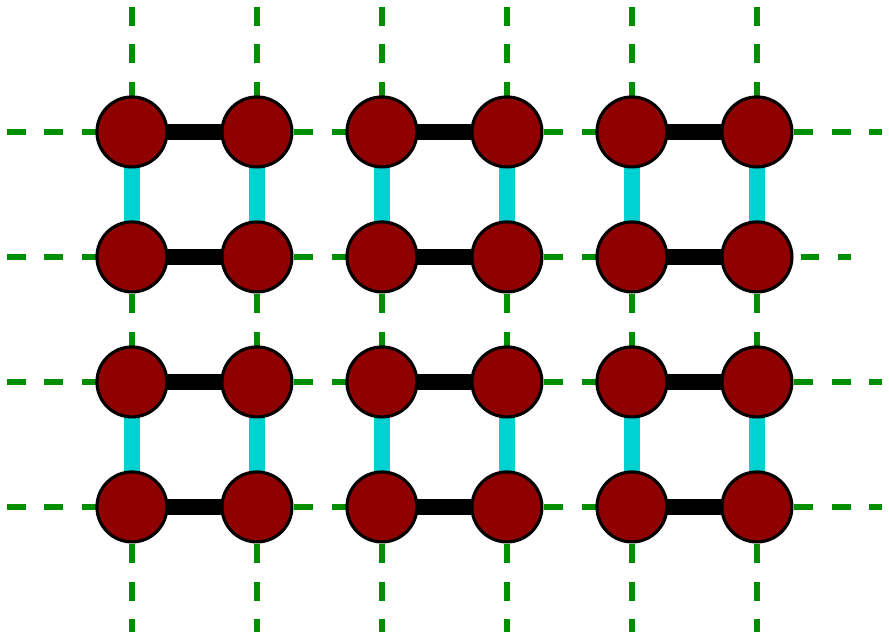}\label{fig:cpt}}\\[1em]
\subfigure[]{\includegraphics[width=\textwidth,angle=90]{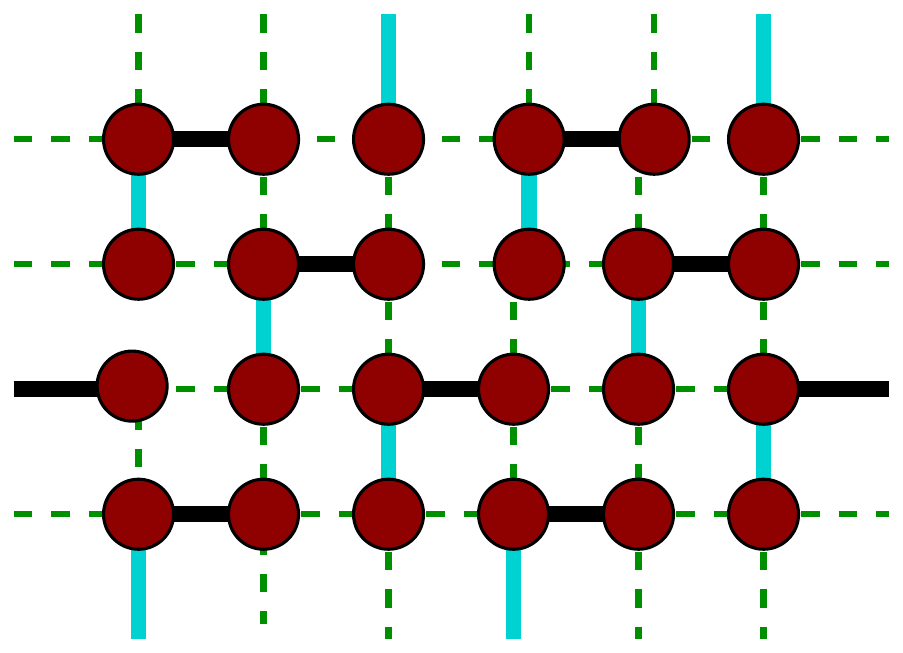}\label{fig:cpt3}}
\end{minipage}
\hfill
\begin{minipage}{0.65\columnwidth}
\subfigure{\includegraphics[width=\textwidth]{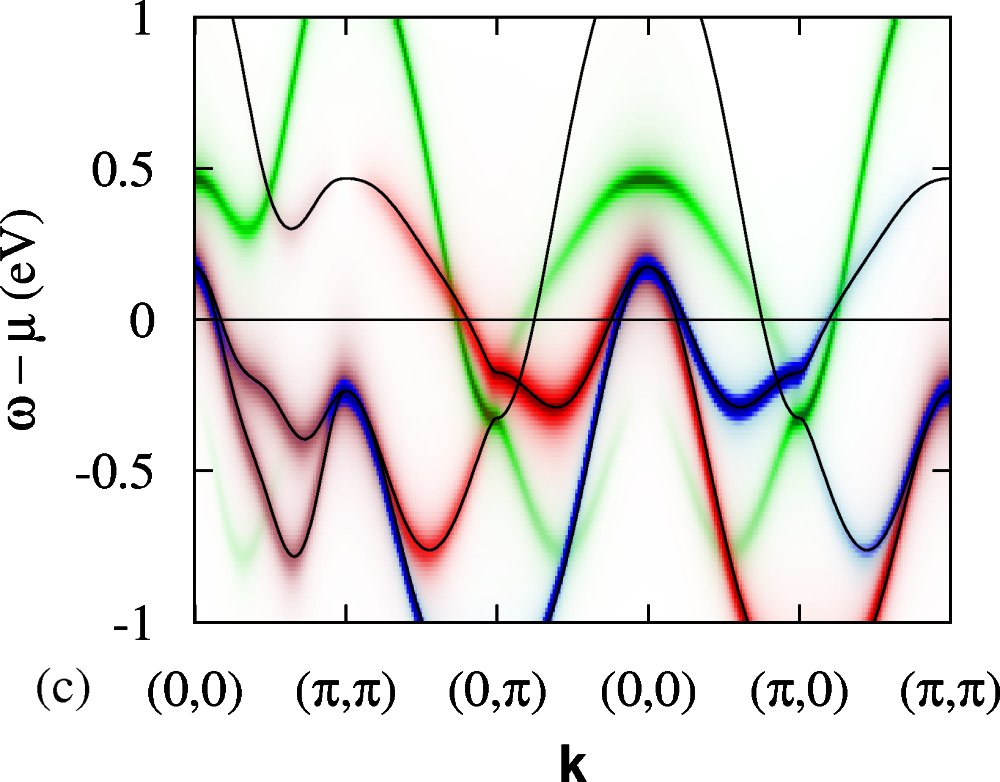}\label{fig:3b_U0}}\\[-1.5em]
\subfigure{\includegraphics[width=\textwidth]{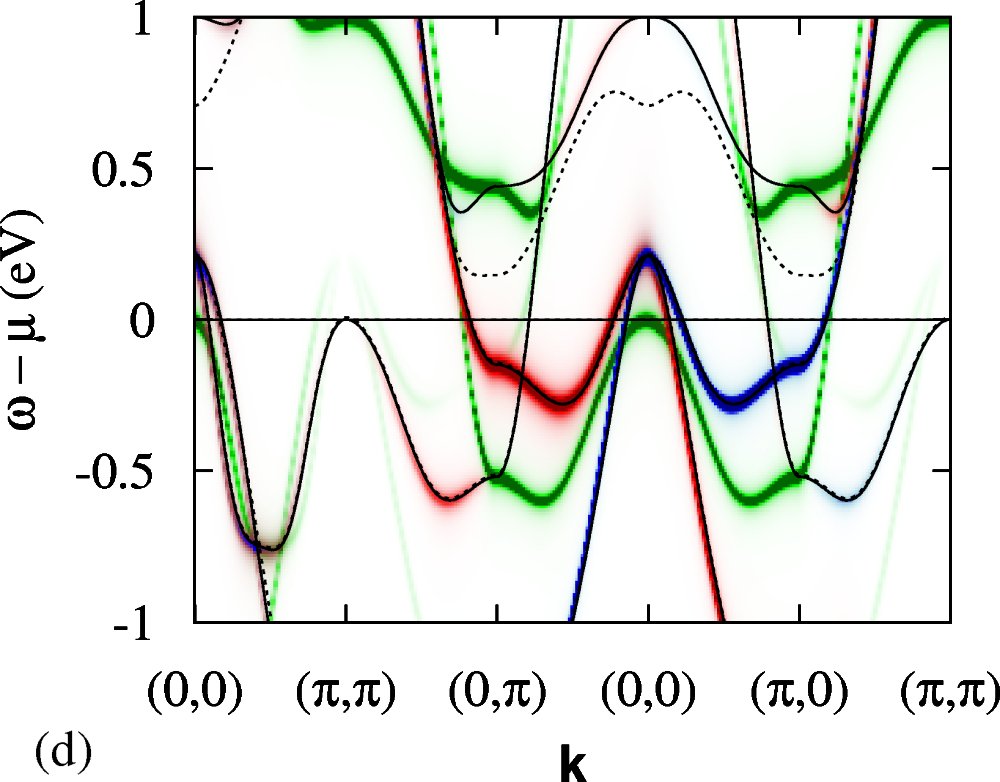}\label{fig:4b_U0}}
\end{minipage}
\caption{(Color online) Schematic illustration of the cluster
  decomposition into (a) four-site and (b) three-site clusters used for the
  three- and four-orbital models. Ground-state
energies and Green's functions of the small cluster -- as connected by
thick solid lines --  are obtained by exact diagonalization. Clusters
are then connected in CPT along the thinner dashed bonds. Within the
cluster, AFM (FM) Heisenberg exchange acts between electrons along the $x$-
\mbox{($y$-)} bond. 
(c) Spectral density $A({\bf k},\omega)$ of the non-interacting
three-orbital model Eq.~(\ref{eq:3b}). Solid lines indicate the
bands in terms of the pseudo-crystal momentum ${\bf \tilde{k}}$, shading the
spectral weight in terms of ``real'' momentum ${\bf k}$, the
difference is that weight with $xy$ character shifts by
$(\pi,\pi)$. [Note that along $(0,0)$-$(\pi,\pi)$, the bands with
dominant $xz$/$yz$ character contain these two orbitals with 
identical weight, even though the $yz$ character - here drawn on top -
dominates the figures.]
(d) $A({\bf k},\omega)$ of the non-interacting
four-orbital model. Dashed lines indicate the results for the model that is obtained by
removing~\cite{PhysRevB.84.235115} the $3z^2-r^2$ orbital from the five-orbital
model of Ref.~\onlinecite{graser_5b}. Solid lines are for the model
used here, where the $x^2-y^2$ orbital is then somewhat removed from
the Fermi level by changing $t_{xx}^{33}$ from $-0.02$ to
$t_{xx}^{33}=0.03$ and $\epsilon_3$ from $-0.22$ to $-0.12$ (notation
as in Ref.~\onlinecite{graser_5b}). 
As in (c), shading indicates the spectral weight for the
``real'' momentum instead of the pseudo-crystal momentum. 
In the online version, red
  refers to $xz$, blue to $yx$, and green to all other orbitals. In all spectra, peaks are broadened by a
    Lorentzian $\delta/((\omega-\omega_0)^2+\delta^2)$ with $\delta
    =0.05$ except for Fig. 1(d), where $\delta =0.025$. All
  energies are in eV. 
\label{fig:model}  
}
\end{figure}

The aim of the paper is to calculate the spectral density $A({\bf
  k},\omega)$ in a phase where (short-range) magnetic correlations
break the fourfold symmetry of the lattice, but without long-range
magnetic order.  The latter requirement prevents us from carrying out
our calculations directly in momentum space, as it has been done for the
paramagnetic and AFM phases.
As an alternative approach, we choose here cluster perturbation
theory.~\cite{Gros:1993p2667,Senechal:2000p2666} 
In this method, the
ground state and one-particle Green's function are evaluated (almost)
exactly (with Lanczos exact diagonalization) for a fully interacting
quantum model on a small cluster, and hoppings between clusters are
treated in perturbation theory [for an illustration see
Figs.~\ref{fig:cpt} and~\ref{fig:cpt3}]. Apart from the limit of small
intercluster hoppings, this approximation also becomes exact in the
opposite limit of vanishing interactions, as can be seen by
considering that it amounts to replacing the self energy of the full
system by that of the small cluster.~\cite{Aic03} Long-range order can
be treated with the related variational cluster approach (VCA),~\cite{Aic03,Dahnken:2004p2409} as it 
has been done for a two-orbital model for
pnictides.~\cite{Daghofer:2008,Yu:2009p2127}

The biggest drawback of the VCA is that correlations are only included exactly
within the small cluster, while longer-range effects are treated at a
mean-field level. For nematic phases with at most short-range order,
this limitation turns into a huge \emph{advantage}: We can break the symmetry
between the $x$ and $y$ directions locally on the small cluster, see
below, but without imposing long-range 
order by a symmetry-breaking field. If the small cluster is, e.g., an AFM
coupled dimer, its groundstate is thus still given by a singlet, i.e., a
superposition of ``up-down'' and ``down-up'', which removes long-range
correlations. 

When using a dimer as
the directly solved cluster, we find instabilities, i.e., poles of the
one-particle Green's function that are on the wrong side of the
chemical potential. While this does not necessarily invalidate the
results (which are in fact similar to the more stable results
described below), it may indicate that the self energy of a
dimer differs too strongly from that of a large two-dimensional system
to provide a reliable approximation. In order to be able to
use three- (four-) site clusters instead, which lead to stable results, we
restrict the Hamiltonian to the four (three) orbitals that contribute most of
the weight at the FS. The results presented here were obtained with the
cluster decompositions shown in Figs.~\ref{fig:cpt}
and~\ref{fig:cpt3}, but equivalent results were found for the
three-orbital model when using a ``brick-wall'' arrangement of $2\times
2$ clusters instead of the ``columns'' in Fig.~\ref{fig:cpt}.  

The momentum-dependent tight-binding Hamiltonian in orbital
space can be written as 
\begin{align}\label{eq:H0k}
H_{\rm TB}(\mathbf{ \tilde{k}}) = \sum_{\mathbf{ \tilde{k}},\sigma,\mu,\nu} 
T^{\mu,\nu}(\mathbf{ \tilde{k}}) 
d^\dagger_{\mathbf{ \tilde{k}},\mu,\sigma} d^{\phantom{\dagger}}_{\mathbf{ \tilde{k}},\nu,\sigma}\;,
\end{align}
where $d^{\phantom{\dagger}}_{\mathbf{ \tilde{k}},\nu,\sigma}$
($d^{\dagger}_{\mathbf{ \tilde{k}},\nu,\sigma}$) annihilates (creates) an
electron with pseudo-crystal momentum ${\bf \tilde{k}}$ and spin $\sigma$ in orbital $\nu$. 
The three-orbital model used here is based on the model
of Ref.~\onlinecite{Daghofer_3orb}, but a few longer-range hoppings were
added to provide a better fit of the bands near the FS, because the
original three-orbital model has magnetic 
instabilities too far from $(\pi,0)$/$(0,\pi)$.~\cite{Brydon:2011bl}
The $T^{\mu,\nu}(\mathbf{ \tilde{k}})$ give the hoppings between
orbitals $\mu$ and $\nu$ and are  
\begin{align}
T^{11/22} &= 2t_{2/1}\cos  k_x +2t_{1/2}\cos  k_y +4t_3 \cos  k_x 
\cos  k_y \nonumber \\
 &\pm 2t_{11}(\cos 2k_x-\cos 2k_y)+4t_{12}\cos 2 k_x \cos 2k_y,  \\
T^{33} &= \Delta_{xy} + 2t_5(\cos  k_x+\cos  k_y) +4t_6\cos  k_x\cos k_y \nonumber\\
       &\quad  + 2t_9(\cos 2k_x+\cos 2k_y) \nonumber  \\
       &\quad + 4t_{10}(\cos 2k_x \cos k_y + \cos k_x \cos 2k_y),\\
T^{12} &= T^{21} = 4t_4\sin  k_x \sin  k_y,  \label{eq:3b}\\
T^{13} &= \bar{T}^{31} = 2it_7\sin  k_x + 4it_8\sin  k_x \cos  k_y,  \\
T^{23} &= \bar{T}^{32} =  2it_7\sin  k_y + 4it_8\sin  k_y \cos  k_x,  
\end{align}
%
where a bar denotes the complex conjugate. Hopping parameters are $t_1=-0.08$, $t_2=0.1825$,
$t_3=0.08375$, $t_4=-0.03$, $t_5=0.15$, $t_6=0.15$, $t_7=-0.12$,
$t_8=-t_7/2=0.06$, $t_{10} = -0.024$, $t_{11} = -0.01$, $t_{12} =
0.0275$, $\Delta_{xy} = 0.75$, $\mu = 0.4745$; Fig.~\ref{fig:model} shows the
uncorrelated tight-binding bands. The four-orbital model was obtained by
removing the $3z^2-r^2$ orbital~\cite{PhysRevB.84.235115} from the
five-orbital model of Ref.~\onlinecite{graser_5b} and slightly changing
onsite energy and third-neighbor hopping of the $x^2-y^2$ orbital to
alleviate the fact that removing the $3z^2-r^2$ orbital moves it too
close to the Fermi level, see Fig.~\ref{fig:4b_U0}. In
principle, hoppings can be extended to three dimensions and
parameters could be fitted to model specific compounds, at least in
the more detailed four-orbital model. The features we aim
to study here -- an anisotropy between the $X$ and $Y$ points --
have been experimentally observed in different compounds, and we are going to see
that both the three- and the four-orbital models lead to similar
results despite their somewhat different dispersions, suggesting that
fine-tuning of the kinetic energy is not crucial.

We use a unit cell with one iron atom to distinguish between momenta
$(\pi,0)$ and $(0,\pi)$, which  would both map to $(\pi,\pi)$ for a two-iron unit cell. 
Due to an internal symmetry of the two-iron
unit cell,~\cite{plee,eschrig_tb} 
it is always possible to use a one-iron unit cell for tight-binding
models restricted to an Fe-As plane. However, the $xz$ and
$yz$ orbitals with
momentum ${\bf k}$ couple to the other orbitals at momentum ${\bf
k}+(\pi,\pi)$. Thus, one writes the tight-binding Hamiltonians in terms
of a pseudo-crystal momentum $\bf{\tilde{k}}$, which is $\bf{\tilde{k}}=\bf{k}$ for
$xz$/$yz$ and $\bf{\tilde{k}}={\bf k}+(\pi,\pi)$ for
  $xy$/$x^2-y^2$/$3z^2-r^2$. 
In real space, such a  notation corresponds to a local gauge
transformation, where replacing, e.g., the $xy_{\bf i}$ orbital at site
${\bf i}=(i_x,i_y)$ by $(-1)^{(i_x+i_y)}xy_{\bf i}$ (and analogously for
$x^2-y^2$ and $3z^2-r^2$) leads to a
translationally invariant Hamiltonian with a one-iron unit cell. For comparison with ARPES
experiments, however, this gauge transformation has to be undone,
which implies that spectral weight at $\bf{\tilde{k}}$ with orbital
character $xy$, $x^2-y^2$ or $3z^2-r^2$ is plotted at
$\bf{k}=\bf{\tilde{k}}+(\pi,\pi)$.~\cite{Tranquada_spinorb,oo_nematic_2011}

In order to study a nematic phase, the four-fold lattice symmetry is
explicitly broken by
introducing a phenomenological Heisenberg interaction, which couples
electrons in all orbitals,
\begin{align}\label{eq:Heisenberg}
H_{\textrm{Heis}} = \pm J_{\textrm{anis}} \sum_{\stackrel{\langle {\bf i},{\bf
    j}\rangle\parallel x/y}{\mu\nu}}
{\bf S}_{{\bf i}\mu}\cdot
{\bf S}_{{\bf j}\nu},
\end{align}
where $\mu$, $\nu$ denote orbitals and $\langle {\bf i},{\bf
  j}\rangle$ nearest-neighbor (NN) bonds. For $J_{\textrm{anis}}>0$, the coupling is
AFM (FM) along the $x$ ($y$) direction. The
electron-spin operators are given by ${\bf S}_{{\bf i}\nu} = \tfrac{1}{2}\sum_{ss'}
d^{\dagger}_{{\bf i}\nu s}{\boldsymbol \sigma}^{\phantom{\dagger}}_{ss'}d^{\phantom{\dagger}}_{{\bf
    i}\nu s'}$, where ${\boldsymbol \sigma} = (\sigma^x,\sigma^y,\sigma^z)$
is the vector of Pauli matrices. These interactions act only within the
small cluster that is solved exactly. 
We are here not
going to investigate the origin of such a breaking of rotational
symmetry, which has been shown to occur in several
models,~\cite{Fang:2008nematic,Xu2,Fernandes_nematic2012,PhysRevB.82.020408} 
but we will study its impact on the system. We find that
when the system is 
close to the spin-density wave, very small values of
$J_{\textrm{anis}}$ trigger highly anisotropic band distortions,
suggesting that short-range correlations, as observed in a
spin-fermion model,~\cite{Shuhua_Pnict2011} indeed favor such a
symmetry breaking.

When onsite interactions~\cite{PhysRevB.18.4945,Oles:1983} are taken into account, the same values of
intra-orbital Coulomb repulsion $U$, inter-orbital repulsion $U'$,
Hund's rule coupling $J$ and pair hopping $J'=J$ were used for all orbitals, along with the standard relation $U'=U-2J$, giving
\begin{equation}\begin{split}  \label{eq:Hcoul}
  H_{\rm int}& =
  U\sum_{{\bf i},\alpha}n_{{\bf i},\alpha,\uparrow}n_{{\bf i},
    \alpha,\downarrow}
  +(U'-J/2)\sum_{{\bf i},
    \alpha < \beta}n_{{\bf i},\alpha}n_{{\bf i},\beta}\\
  &\quad -2J\sum_{{\bf i},\alpha < \beta}{\bf S}_{\bf{i},\alpha}\cdot{\bf S}_{\bf{i},\beta}\\
  &\quad +J\sum_{{\bf i},\alpha < \beta}(d^{\dagger}_{{\bf i},\alpha,\uparrow}
  d^{\dagger}_{{\bf i},\alpha,\downarrow}d^{\phantom{\dagger}}_{{\bf i},\beta,\downarrow}
  d^{\phantom{\dagger}}_{{\bf i},\beta,\uparrow}+h.c.),
\end{split}\end{equation}
where $\alpha,\beta$ denote the orbital and ${\bf S}_{{\bf i},\alpha}$
($n_{{\bf i},\alpha}$) is the spin (electronic density) in orbital $\alpha$ at site
${\bf i}$. 
While the parameters relating to the $xy$ and $x^2-y^2$ orbitals can in principle be slightly different from each other and the $xz$/$yz$ doublet, symmetric interactions were chosen for simplicity.

\section{Results}\label{sec:results}

\subsection{Band anisotropy in the three- and four-orbital models}\label{sec:results_U0}

\begin{figure}
\subfigure{\includegraphics[width=0.27\textwidth]{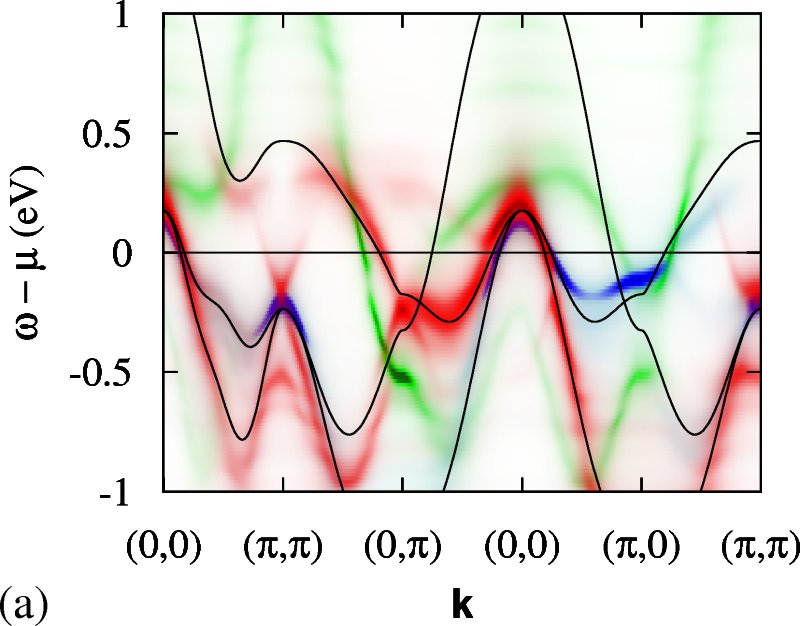}\label{fig:3b_U0_Janis}}
\subfigure{\includegraphics[width=0.2\textwidth]{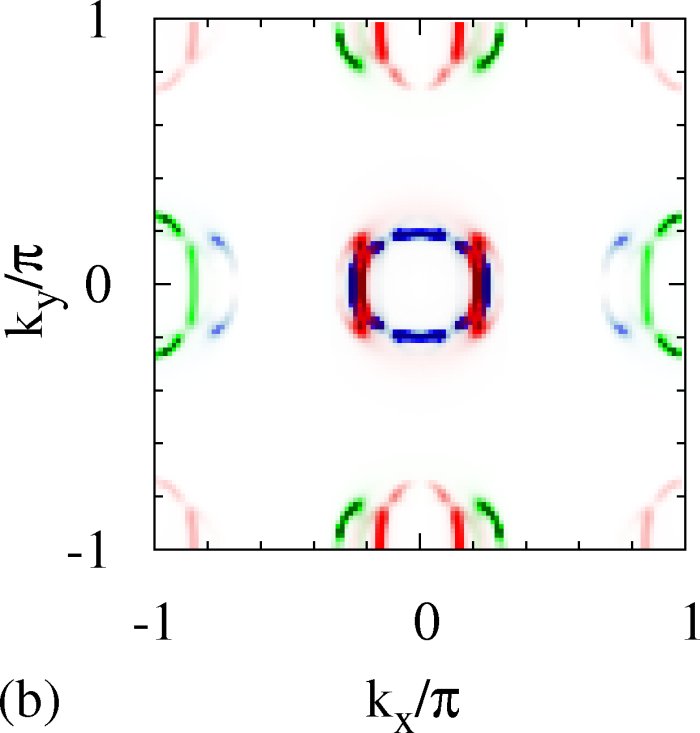}\label{fig:fs_3b_U0_Janis}}\\
\caption{(Color online) (a) Spectral density $A({\bf k},\omega)$ and (b) Fermi
  surface of the three-orbital model (four-site cluster) with parameters
  as given for Fig.~\ref{fig:3b_U0} and an explicit symmetry-breaking
  $J_{\textrm{anis}}=0.5\;\textrm{eV}$, see
  Eq.~(\ref{eq:Heisenberg}), but without Coulomb repulsion and Hund's rule
  coupling. Shadings are for ``real'' momentum, lines indicate the
  non-interacting model in pseudo-crystal momentum ${\bf \tilde{k}}$. 
\label{fig:Ak_Janis_U0_3b}}
\end{figure}

In order to study the effects of phenomenological
  short range magnetic correlations, 
the Hamiltonian given by Eqs.~(\ref{eq:H0k}) and
(\ref{eq:Heisenberg}), was initially treated with the VCA on a 
four-site cluster, with AFM interactions along $x$ and FM ones along
$y$ but without onsite Coulomb and Hund interactions. A fictitious
chemical potential was optimized as a variational 
parameter, but did not have a large impact on the results. No
tendencies towards long-range order were found, which agrees with
expectations: Since the AFM Heisenberg interaction only acts within
the cluster, it favors a total cluster spin of $S_{\textrm{tot}}=0$.
In the large system, consisting of many noninteracting clusters,
there is no magnetic order.  Rather large 
$J_{\textrm{anis}}\gtrsim 0.3\;\textrm{eV}$ has to be
chosen to induce appreciable signatures of the
anisotropy, which is a very large energy scale compared to the other
parameters of the Hamiltonian. The reason is that the non-interacting
model with four electrons per site does not contain any net unpaired
spins that can directly be coupled by a Heisenberg interaction; the 
interaction first needs to be strong enough to induce a local spin.

The spectral density for $J_{\textrm{anis}}=0.5\;\textrm{eV}$ is shown
in Fig.~\ref{fig:3b_U0_Janis}. Apart from the fact that the Heisenberg
interactions make the spectrum more incoherent, the bands are most
strongly modified near
$X=(\pi,0)$, which corresponds to the ordering vector that would be
favored by the NN AFM interaction along $x$. One clearly sees that the $yz$ states around $X$ are moved
to higher energies, while the $xz$ states
at $Y=(0,\pi)$ are shifted to slightly lower energies in agreement with experiments. The energy
shifts are momentum dependent: While the differences between $X$ and
$Y$ are large, changes around $\Gamma=(0,0)$ are far less
pronounced. The corresponding orbital-resolved FS can be seen in
Fig.~\ref{fig:fs_3b_U0_Janis}. Like the spectral density, it shows
some features that are similar to those resulting from band folding in
a $(\pi,0)$ SDW; for example, the $xz$ electron pocket at $Y$ has a
``mirror pocket'' at $M=(\pi,\pi)=Y+(\pi,0)$. However, the FS
is still qualitatively different from the FS found in the
long-range ordered SDW, where folding leads to additional features and
largely suppresses the $yz$ weight,~\cite{orb_pol_FS} which dominates
the hole pockets here. Such differences
related to long-range order are consistent with ARPES experiments in
NaFeAs,~\cite{ARPES_NaFeAs11} see also the discussion in
Sec.~\ref{sec:results_U}. 

\begin{figure}
\subfigure{\includegraphics[width=0.35\textwidth]{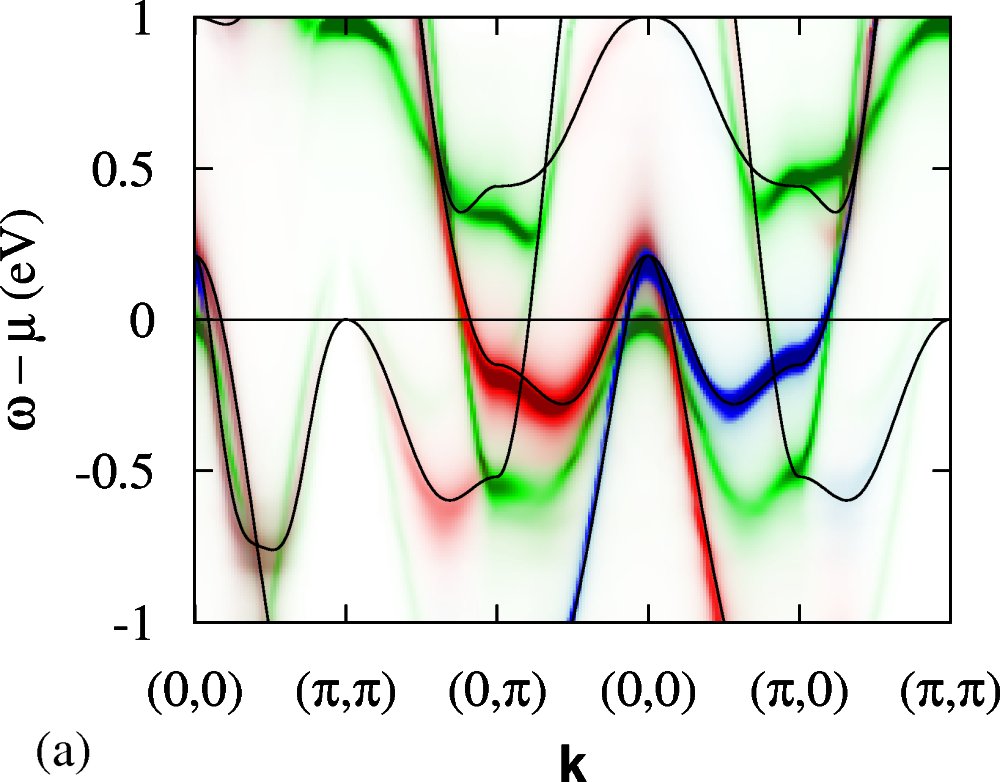}\label{fig:4b_U0_J02}}\\[-0.5em]
\subfigure{\includegraphics[width=0.35\textwidth]{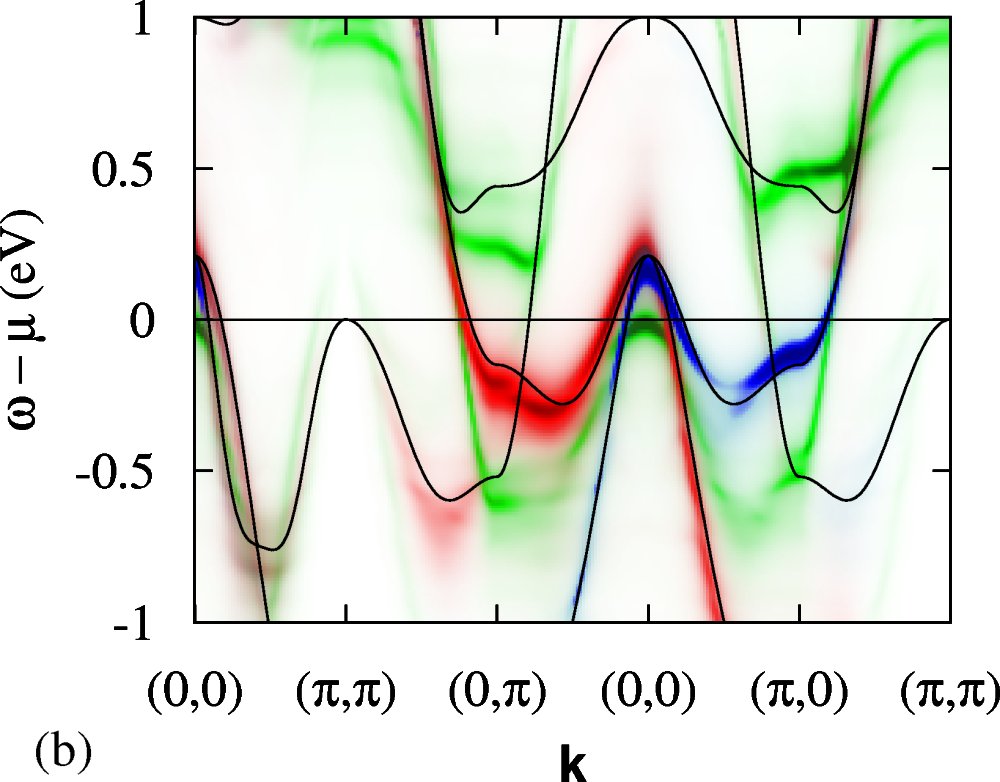}\label{fig:4b_U0_J03}}\\[-0.5em]
\subfigure{\includegraphics[width=0.35\textwidth]{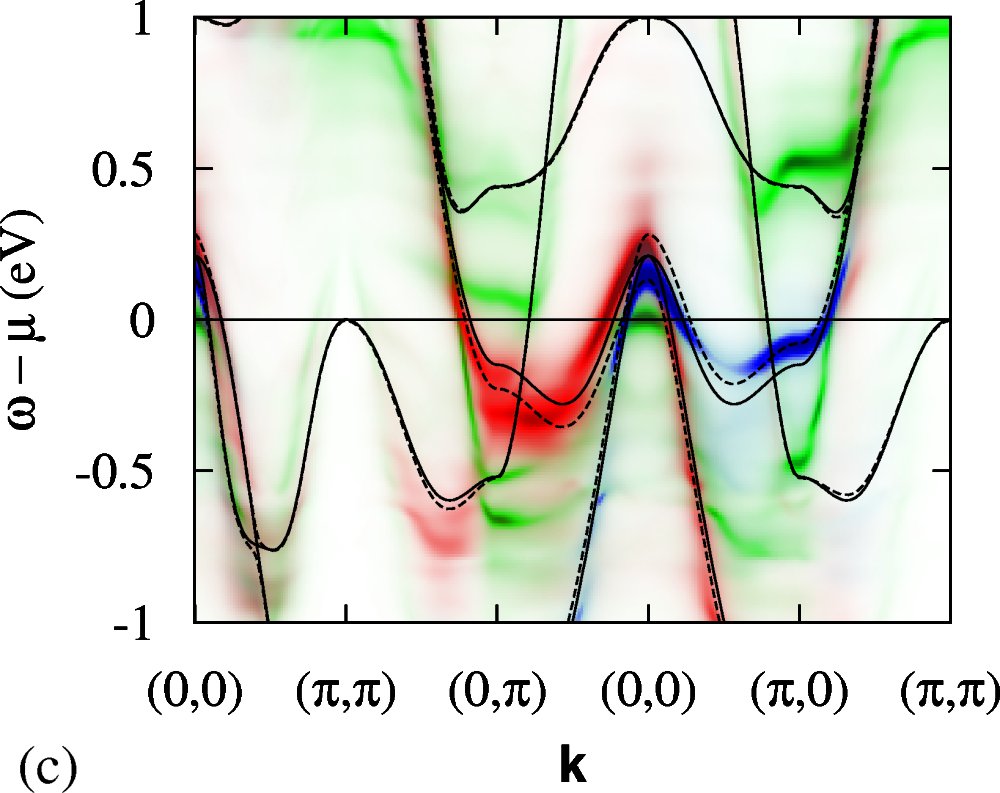}\label{fig:4b_U0_J04}}
\caption{(Color online) Spectral density $A({\bf k},\omega)$  of the
  four-orbital model (three-site cluster), see Fig.~\ref{fig:4b_U0}, and an increasing explicit
  symmetry-breaking term Eq.~(\ref{eq:Heisenberg}) of (a)
  $J_{\textrm{anis}}=0.2\;\textrm{eV}$, (b)
  $J_{\textrm{anis}}=0.3\;\textrm{eV}$, and (c)
  $J_{\textrm{anis}}=0.4\;\textrm{eV}$.  Coulomb repulsion and Hund's rule
  coupling are not included. Shadings are for ``real'' momentum, solid lines indicate the
  non-interacting model in pseudo-crystal momentum ${\bf
    \tilde{k}}$. In (c), dashed lines are for a non-interacting model
  with an energy difference $\Delta =0.15\;\textrm{eV}$ between the $xz$ and $yz$ orbitals, which was
  fitted to approximately reproduce the difference between the $X$ and
  $Y$ points.
\label{fig:Ak_Janis_U0_4b}}
\end{figure}

The same behavior as in the three-orbital model is seen for the
four-orbital case, see Fig.~\ref{fig:Ak_Janis_U0_4b}, where $A({\bf
  k},\omega)$ is shown for increasing
$J_{\textrm{anis}}=0.2,\;0.3,\;0.4\;\textrm{eV}$. In the last case,
the splitting between the states at $X$ and $Y$ is $\approx
150\;\textrm{meV}$. Taking into account that the overall band width has
to be renormalized by a factor of 2-3, this is consistent with the order of
magnitude of the $60\;\textrm{meV}$ splitting reported for
Ba(Fe$_{1-x}$Co$_x$)$_2$As$_2$.~\cite{Yi:PNAS2011} This can be
compared to an explicit orbital splitting, similar to the mechanism proposed in
Ref.~\onlinecite{oo_nematic_2011}. The splitting can be written as 
$\Delta=(n_{yz}-n_{xz})/2$, where $n_{xz}$ ($n_{yz}$) is the
density in the $xz$ ($yz$) orbital, and was set to $\Delta
=0.15\;\textrm{eV}$, which approximately reproduces the energy
difference between the $X$ and $Y$ points indicated by the dashed lines in Fig.~\ref{fig:4b_U0_J04}. A 
momentum-independent splitting large enough to reproduce the energy
differences between the $X$ and $Y$ points substantially distorts the features
near the $\Gamma$ point as well. While the unoccupied states above the
chemical potential at $\Gamma$ are not easily accessible in ARPES, available data
on the bands defining the hole pockets appear more consistent with
the slighter changes caused by momentum-dependent shifts of the
nematic scenario, especially for cases with a large splitting between the features near
$X$/$Y$, where the bands near $\Gamma$ would be very strongly distorted by
a rigid shift.~\cite{Yi:PNAS2011,ARPES_NaFeAs11,He:2010pNaFeAs,ZhangARPES_NaFeAs_2011}

\begin{figure}
\subfigure[]{\includegraphics[height=0.61\columnwidth,trim=60 0 75 0,clip]{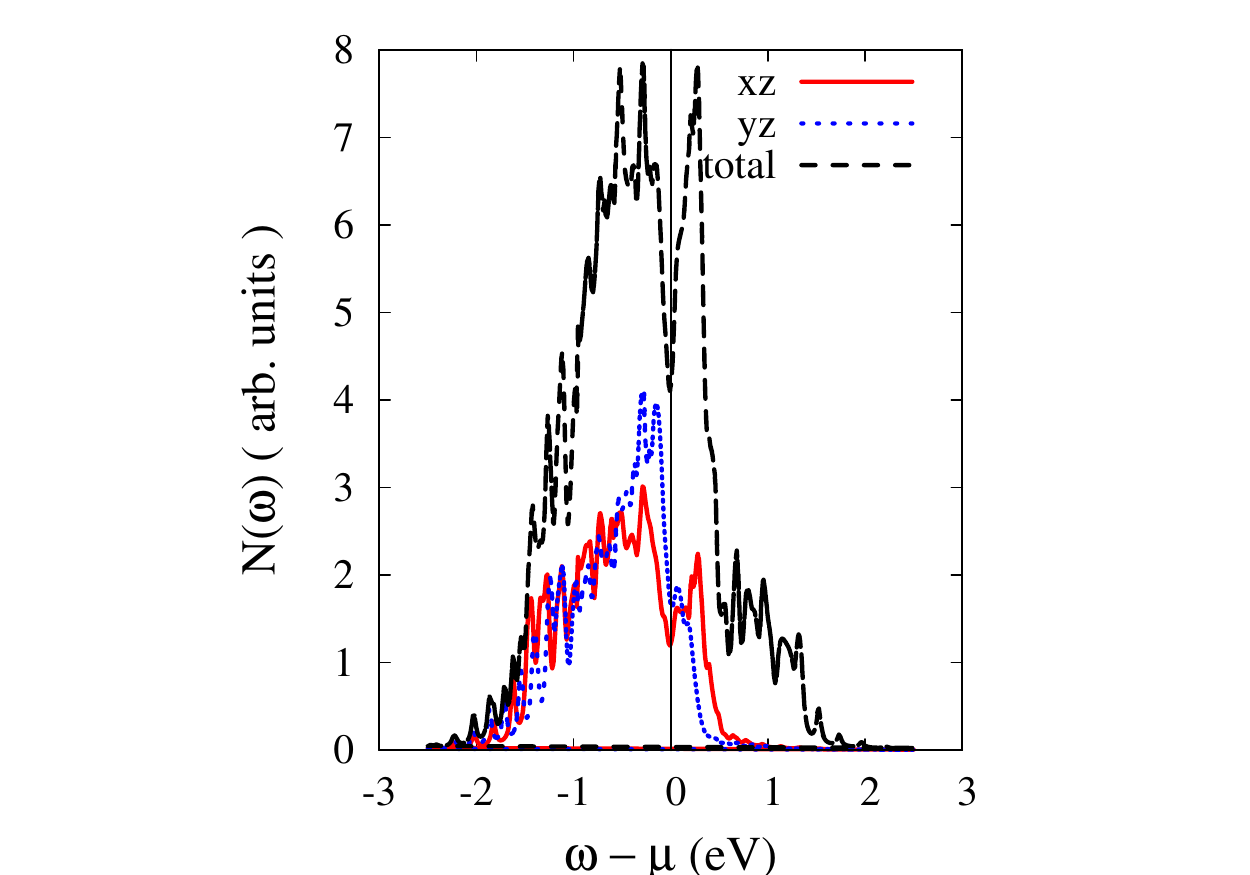}\label{fig:dos_3b}}
\subfigure[]{\includegraphics[height=0.61\columnwidth,trim=105 0 75 0,clip]{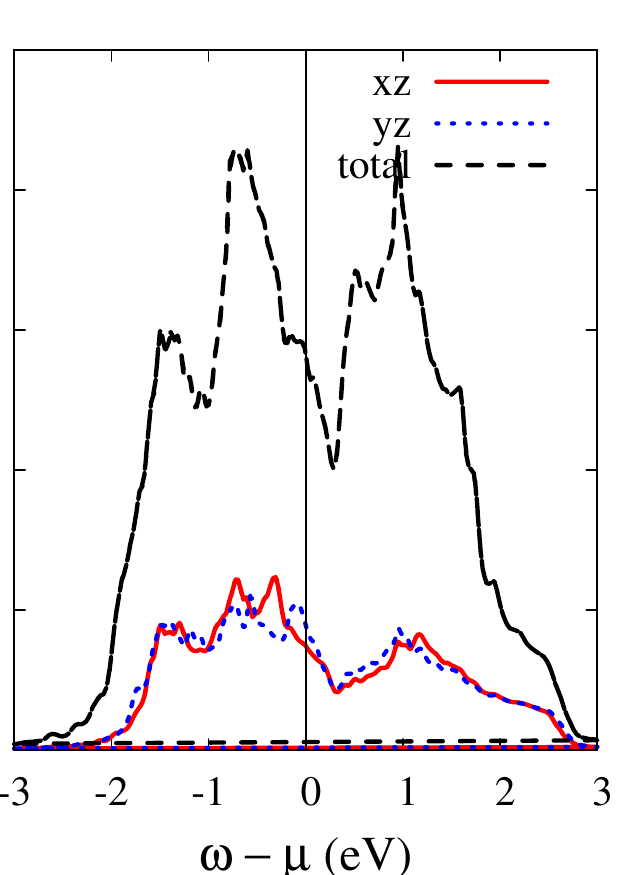}\label{fig:dos_4b}}\\
\caption{(Color online) Density of states for (a) the three-orbital model
  with $J_{\textrm{anis}}=0.5\;\textrm{eV}$ and (b) the four-orbital
  model and
  $J_{\textrm{anis}}=0.4\;\textrm{eV}$. $U=J_{\textrm{Hund}}=0$ in
  both cases. 
\label{fig:dos_Janis}}
\end{figure}

Total orbital densities do not turn out to be a reliable way to
characterize the impact of the nematic order on states near the Fermi
energy. Densities in the $xz$ and $yz$ orbitals differ only slightly
in the four-orbital model with $n_{xz}-n_{yz}\approx 0.02$ for
$J_{\textrm{anis}}=0.4\;\textrm{eV}$. This value is not strongly
affected by $4\%$ hole or electron doping, in contrast to a proposed
sign change for hole doping~\cite{Fernandes_nematic2012} and it is broadly
consistent with the small orbital polarizations found in mean-field
analyses for the SDW state.~\cite{Daghofer_3orb,orb_pol_FS} In the
three-orbital model, the orbital polarization is even
\emph{opposite} with $n_{xz}-n_{yz}\approx -0.1$, because spectral weight
with $yz$ character is transferred below the Fermi level [see the
density of states shown in Fig.~\ref{fig:dos_3b}], and in
contrast to the four-orbital model [see Fig.~\ref{fig:dos_4b}], this weight is not balanced by $xz$ states
further away from $\mu$. Nevertheless, the band reconstruction near
the Fermi level and the band anisotropy are very similar in the two
models. AFM correlations along $x$ always bring the $yz$
states around $X$ closer to the Fermi level, even when the
total orbital densities satisfy $n_{yz}>n_{xz}$, in
  contrast to a naive expectation that the $yz$ bands should be
  lowered in energy in this case. When
  onsite interactions bring the three-orbital 
  model closer to the SDW transition (see Sec.~\ref{sec:results_U}
  below), the orbital densities become
  almost equal with $n_{xz}-n_{yz}\approx -0.012$ for
  $U=1\;\textrm{eV}$. Total
orbital densities can determine magnetic properties via the 
Goodenough-Kanamori rules in Mott insulators, which do not have a
FS. On the other hand, since the pnictides are more metallic 
with strongly hybridized orbitals, a more consistent and clearer
picture can here be obtained if one concentrates on spectral weight
near the Fermi level as will be discussed below. 

\subsection{Impact of onsite Coulomb interaction}\label{sec:results_U}

\begin{figure}
\subfigure{\includegraphics[width=0.3\textwidth]{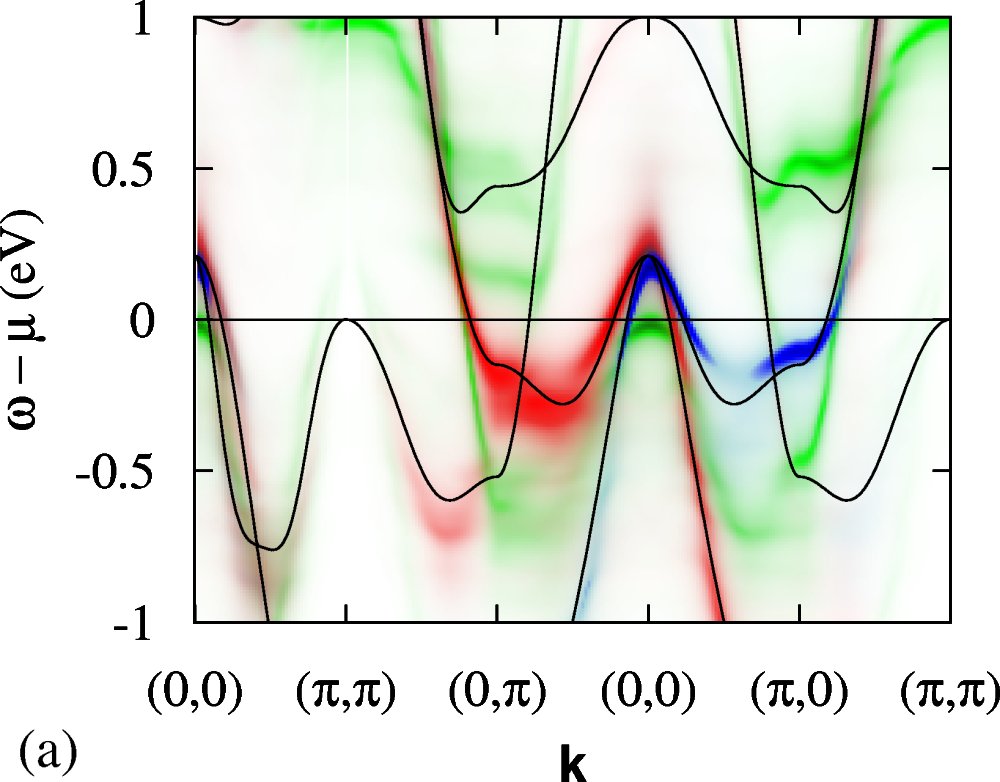}\label{fig:4b_U03_Janis}}\\
\subfigure{\includegraphics[width=0.3\textwidth]{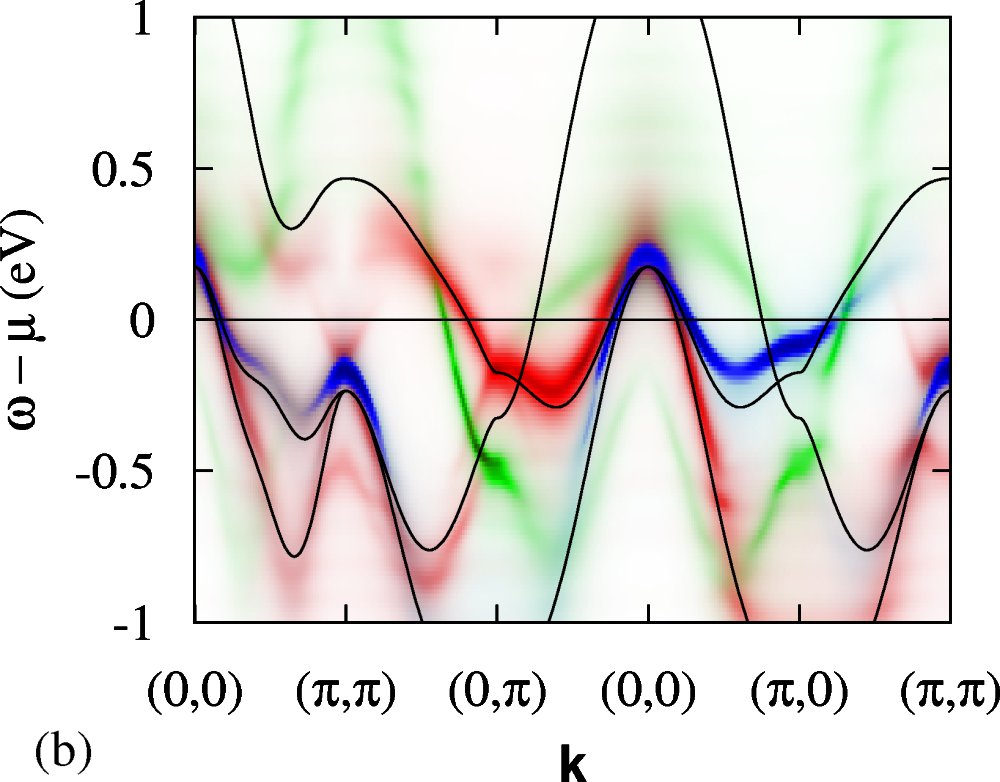}\label{fig:3b_U06_Janis}}\\
\subfigure{\includegraphics[width=0.3\textwidth]{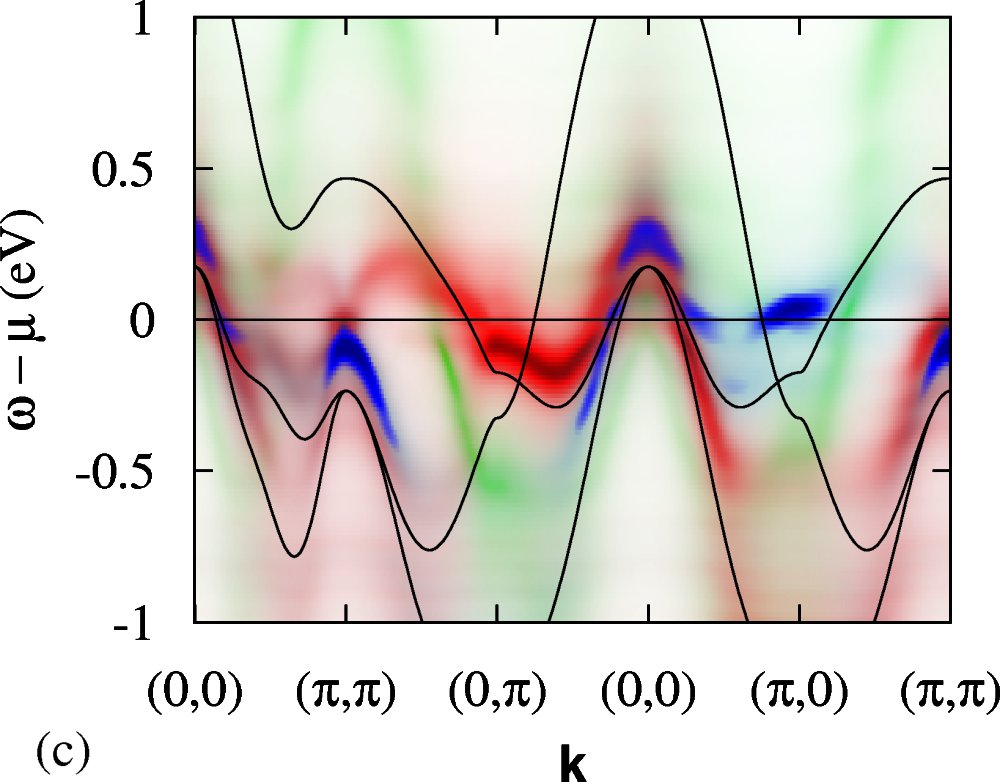}\label{fig:3b_U12_Janis}}
\subfigure{\includegraphics[width=0.17\textwidth]{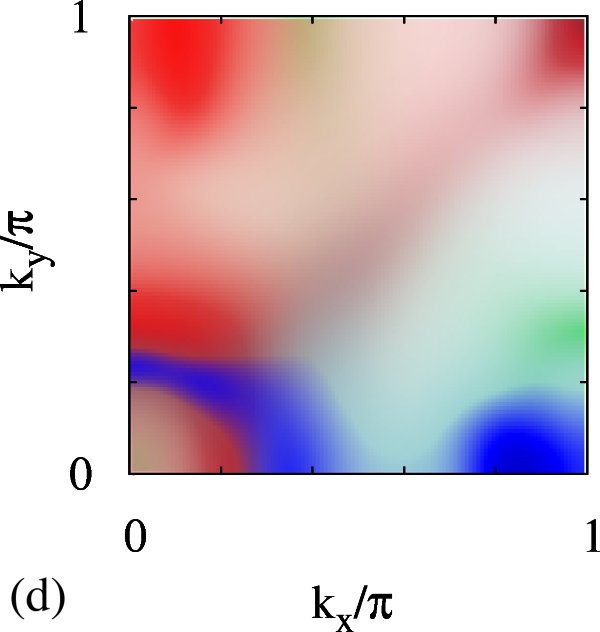}\label{fig:fs_3b_U12_Janis}}\\
\caption{(Color online) 
Spectral density with anisotropic short-range magnetic order and
onsite interactions. (a) For the four-orbital model and
$U=0.3\;\textrm{eV}$, $J_{\textrm{Hund}}=0.075\;\textrm{eV}$  and
$J_{\textrm{anis}}=0.2\;\textrm{eV}$.
(b) For the three-orbital model with $U=0.6\;\textrm{eV}$ 
($J_{\textrm{Hund}}=0.15\;\textrm{eV}$) and
$J_{\textrm{anis}}=0.2\;\textrm{eV}$ and (c) $U=1.02\;\textrm{eV}$ 
($J_{\textrm{Hund}}=0.255\;\textrm{eV}$) and
$J_{\textrm{anis}}=0.015\;\textrm{eV}$. For these last values of $U$
and $J_{\textrm{Hund}}$, the three-orbital model is very close to the
SDW. (d) FS corresponding to the parameters in (c), it
captures spectral weight within $1\;\textrm{meV}$ of the Fermi level;
broadening of the spectral weight is consistent with (c).  
Onsite
interactions are lower for the four-orbital model, because it is at half
filling and Hund's rule thus, moves it closer to a Mott transition,
while it partly compensates $U$ away from half filling, as in the
three-orbital model. Shadings are for ``real'' momentum;
lines indicate the non-interacting model in pseudo-crystal momentum ${\bf
\tilde{k}}$. 
\label{fig:Ak_Janis_withU}}
\end{figure}

In this subsection the impact of onsite interactions will be
investigated. The full Eq.~(\ref{eq:Hcoul}) including spin-flip and pair-hopping terms can
easily be included in the VCA. Interaction strengths were chosen 
below the critical values for the onset of long-range order because
we want to focus on short-range correlations 
here. As can be seen in Fig.~\ref{fig:Ak_Janis_withU} for the two models considered here, lower values
of $J_{\textrm{anis}} \approx 0.2$ is now
sufficient to induce substantial asymmetries, in contrast to the
larger $J_{\textrm{anis}}\approx 0.4$ to $0.5\;\textrm{eV}$ needed for
the noninteracting models. Onsite interactions favor local
magnetic moments, even in the absence of long-range order, that can
then be coupled even by weaker $J_{\textrm{anis}}$. 

Finally, we study
the three-orbital model very close to the SDW transition by setting
$U=1.02\;\textrm{eV}$. In a mean-field treatment as used in
Ref.~\onlinecite{Daghofer_3orb}, one finds an SDW with long-range
magnetic order, but the optimal VCA solution does not yet show
long-range order due to the presence of quantum fluctuations. However,
the system is so close to a magnetically ordered state that very small 
$J_{\textrm{anis}}\approx 0.015\;\textrm{eV} = 15\;\textrm{meV}$ already introduces strong
short-range order and corresponding band anisotropies. Several occupied low-energy
bands  [e.g. between $\Gamma=(0,0)$ and $Y=(0,\pi)$ as well as around $M=(\pi,\pi)$] in the spectral density, which is shown in
Fig.~\ref{fig:3b_U12_Janis}, have energies reduced by
a factor of $\approx 2$, consistent with the renormalization factor
$\approx 2$ - $3$ needed
to reconcile density-functional bands with ARPES. Bands above the
Fermi level do not have reduced widths. This 
asymmetric impact of correlations is in agreement
with dynamical mean-field studies.~\cite{Aichhorn:2009gd} In addition
to the renormalization, $J_{\textrm{anis}}= 15\;\textrm{meV}$
induces an energy splitting of $\approx 70\;\textrm{meV}$ between the
$X$ and $Y$ points. In fact, the band at $X$ has moved slightly \emph{above 
the chemical potential}, as expected for the SDW phase. The fact, that this happens even in the
absence of long-range order, is in excellent agreement with recent
ARPES data for NaFeAs, where it was likewise found that the overall
band positions at $X$ and $Y$ nearly reach their ``SDW values'' 
above the N\'eel temperature.~\cite{ARPES_NaFeAs11} 

Nevertheless, the
corresponding FS, see Fig.~\ref{fig:fs_3b_U12_Janis}, clearly shows
important differences to that of the SDW state: As the $yz$ states
cross the chemical potential here with a rather low Fermi velocity
(leading to an elongation of the hole pocket at $\Gamma$ along the $x$-direction),
they contribute substantial weight to the FS. In fact, both of the strong
features along the $\Gamma$-$X$ line are of $yz$ character. In the SDW
phase, in contrast, the $yz$ orbital dominates the AFM order parameter
and is thus mostly gapped out.~\cite{orb_pol_FS} Related effects have
likewise been observed in ARPES, where these $yz$ bands open gaps at the
N\'eel temperature.~\cite{Shimojima:2010p2390,ARPES_NaFeAs11}

\section{Summary and Conclusions}\label{sec:conclusions}

The variational cluster approach was used to study the spectral density
of a nematic phase in three- and four-orbital models for iron-based
superconductors. We found that the method is well suited for problems
involving short-range correlations without long-range (magnetic)
order. The correlations considered in this study were extremely short-range, going
only over NN sites, the minimum to break rotational invariance. While this is a somewhat
extreme scenario, it has been argued that magnetic correlations that are effective only on a very 
short range lead to the linear
temperature dependence of the magnetic susceptibility at high
temperatures.~\cite{Klingeler:2010p2503} Nuclear quadrupole
resonance~\cite{NQR_IFW} indicates that there are As ions seeing different
electronic surroundings in the ``underdoped regime'', which would be in
agreement with the present scenario of As ions involved in ``magnetic''
vs. ``non-magnetic'' bonds.

When symmetry between the $x$- and $y$ directions is broken by a
phenomenological magnetic interaction that is AFM in the $x$ direction,
the bands with $yz$ character around momentum $X=(\pi,0)$ move to
higher energies, i.e., closer to the Fermi level. This is in agreement with
ARPES on detwinned samples above the magnetic transition
temperature, in both the ``122'' compound 
Ba(Fe$_{1-x}$Co$_x$)$_2$As$_2$,~\cite{Yi:PNAS2011} and the ``111''
compound NaFeAs.~\cite{ARPES_NaFeAs11,ZhangARPES_NaFeAs_2011} The
latter is not expected to have surface 
states\cite{PhysRevB.82.184518} that might complicate the analysis of
ARPES in 122 compounds.\cite{PhysRevLett.106.027002}
The changes in the band
structure due to the nematic order not only depend on the orbital, but
also on momentum. Changes around the $\Gamma=(0,0)$ point are far less
pronounced than differences between $X$ and $Y$, again in
agreement with ARPES.~\cite{ARPES_NaFeAs11,He:2010pNaFeAs,ZhangARPES_NaFeAs_2011} Total orbital
densities and their difference are model dependent and not a reliable
predictor of reconstructions of low-energy states. However, the orbital-resolved
spectral weight and the bands near the Fermi level
are affected in the same way both in a three- and a four-orbital model,
with and without onsite interactions indicating that
  they are more universal and less dependent on details of the model Hamiltonian. In agreement with previous
findings on the orbital polarization of the FS in the SDW
phase~\cite{orb_pol_FS} and on transport
properties,~\cite{PhysRevB.84.132505,Shuhua_Pnict2011} this suggests
that total (orbital) densities are here less 
important than in Mott insulators, as the metallic character of the
pnictides makes states near the Fermi level far more important
than those further away.

When onsite interactions are strong enough to bring the system close
to the SDW transition, very small anisotropic couplings can deform the
bands until their broad features resemble bands in the SDW regime,
i.e., bands are renormalized by a factor of $\approx 2$ and the $yz$
states at $X$ move above the chemical potential, as 
seen in ARPES on NaFeAs just above the N\'eel temperature.~\cite{ARPES_NaFeAs11} Nevertheless, the Fermi
surface still differs from that of a state with full
long-range magnetic order, where the $yz$ states are
mostly gapped out,~\cite{Shimojima:2010p2390,orb_pol_FS} again in agreement with
ARPES.~\cite{ARPES_NaFeAs11}

\begin{acknowledgments}
This research was sponsored by the Deutsche
Forschungsgemeinschaft (DFG) under the Emmy-Noether program, the NSF
grant DMR-1104386, and the
Division of Materials Science and Engineering, Office of Basic Energy Sciences,
U.S. DOE. We thank Philip Brydon and Jeroen van den Brink for helpful discussion.
\end{acknowledgments}

%

\end{document}